\documentstyle[aps,preprint]{revtex}
\tightenlines
\firstfigfalse

\def\be{\begin{equation}}
\def\ee{\end{equation}}
\def\ba{\begin{eqnarray}}
\def\ea{\end{eqnarray}}
\def\half{{1 \over 2}}

\def\x{{\bf x}}
\def\e{\mbox{e}}
\def\Re{\mbox {Re}}

\def\phii{\phi_{i,n}}
\def\phif{\phi_{f,n}}
\def\tJ{\tilde{J}}
\def\cap{{\cal C}}
\def\ind{{\cal L}}
\def\const{{\rm const}}
\def\Neff{N_{\rm eff}}
\input epsf
\begin{document}
\preprint{quant-ph/0202125}
\date{February 2002; revised June 2002}
\title{Decoherence by a nonlinear environment: canonical vs. microcanonical
case}
\author{S. Khlebnikov and G. Sadiek\footnote{
On leave of absence from Physics Department, Ain Shams University, 
Cairo, Egypt.}
}
\address{
Department of Physics, Purdue University, West Lafayette, IN
47907, USA}
\maketitle
\begin{abstract}
We compare decoherence induced in a simple quantum system (qubit) for two 
different initial states of the environment: canonical (fixed temperature) and
microcanonical (fixed energy), for the general case of a fully interacting
oscillator environment. We find that even a relatively compact oscillator 
bath (with the effective number of degrees of freedom of order 10), initially
in a microcanonical state, will typically cause
decoherence almost indistinguishable from that by a macroscopic, thermal
environment, except possibly at singularities of the environment's specific 
heat (critical points). In the latter case, the precise magnitude 
of the difference between the canonical and microcanonical results depends on 
the critical behavior of the dissipative coefficient, characterizing 
the interaction of the qubit with the environment. 
\end{abstract}
\pacs{PACS numbers: 03.65.Yz}

\section{Introduction}
Recent years have seen significant experimental advances in manipulation
of quantum states in a variety of physical systems 
\cite{Nakamura&al,suny,delft-mit,purdue,ibm-stanford}. 
In addition to the 
intrinsic interest that these experiments have with regard to the fundamentals 
of quantum mechanics, they suggest that a high degree of control and coherence
in simple quantum systems can be achieved, perhaps eventually sufficient to 
implement a useful quantum computation \cite{Deutsch} in an assembly of such 
individual units (qubits).

Unlike a classical computer, in which the only source of errors is uncontrolled
transitions between the states, a quantum computation is sensitive also to
random changes in {\em phases} of the basis states. Such changes 
occur due to interaction of the qubit with the environment. They are referred
to as decoherence, and the time scale over which the phase will drift by an amount
of order one is referred to as decoherence time $t_d$. 
It is advantageous
to make $t_d$ or, more precisely, the ratio $t_d/t_s$, where $t_s$ is the switching
time, as large as possible.

In many cases, the basic operations on qubits (quantum gates) can be approximated 
as evolution of certain two-level systems under an external influence (a pulse of
voltage, current, etc.). In this paper, we concentrate on cases when
the environment is comprised by interacting oscillators, described for brevity
by a single real scalar field $\phi$, and the interaction of the
two-level system with the environment is linear, with the Hamiltonian of the 
form\footnote{Bilinear and higher-order couplings in $\phi$ can be studied in
a similar way.}
\be
H_J(t)  = - \int d^3 x J(\x, t) \phi(\x, t) \; .
\label{Hint}
\ee
In addition to $H_J$, the Hamiltonian of the environment contains the
free Hamiltonian $H_0$ and a Hamiltonian $H_{\rm int}$, the latter describing
nonlinear interactions among the oscillators themselves. Our method is 
sufficiently general to include fully interacting oscillator environments, 
i.e. any reasonable form of $H_{\rm int}$.
One motivation for doing so is the possibility to consider environments that
are near phase transitions (or the vestiges of such in finite systems).

The ``current'' $J(\x, t)$ depends on the state of the two-level system and represents
its switching history.\footnote{We use word ``switching'' to denote a controlled
transition between states of a qubit, and not a switching on and off the interaction
with the bath. The latter interaction in general cannot be switched off at will;
see, however, remarks on the swap gate in the next paragraph.}
This current, of course, does not have to be the usual electric current, although
it may coincide with it in some specific cases of interactions. 
For example, in a persistent current qubit \cite{percur}, 
where the basis states differ by the value of the electric current, $J$
can indeed be interpreted as the current density, and $\phi$ as a component 
of the electromagnetic field.

An important example, where (\ref{Hint}) applies but $J$ is unrelated to
electric current, is the swap
gate based on two coupled quantum dots \cite{Loss&DiVincenzo}. Key features
of this gate are as follows. There is an electronic
spin 1/2 associated with each dot.
These spins are coupled to each other through an exchange interaction, and
the interaction Hamiltonian is given by the energy of the singlet-triplet splitting. 
The exchange interaction can be switched on and off by varying the potential barrier
between the dots; the latter is controlled by gate voltage $v$, which can be 
viewed as a sum of some average voltage ${\bar v}$ and a fluctuation
$\delta v$. Then, the singlet-triplet splitting energy can be written as
\be
E_{T-S}(v) \approx E_{T-S}({\bar v}) 
+ \frac{\partial E_{T-S}}{\partial v} \delta v \; .
\label{Hdot}
\ee
The first term contributes to the Hamiltonian of the qubit, while the second term
describes the interaction of the qubit with the environment. We see that
this second term is precisely of the form (\ref{Hint}), with
$-\partial E_{T-S}/\partial v$ playing the role of the ``current'', and
$\delta v$ the role of the environment. For the case when fluctuations of $v$ are 
the usual thermal (Nyquist) fluctuations, decoherence induced by $\delta v$ was
considered previously in ref. \cite{Hu&DasSarma}.

The splitting energy $E_{T-S}$ is substantial only during
a pulse of voltage that temporarily lowers the potential barrier. Thus, the duration
of the pulse is the switching time of the gate.
From (\ref{Hdot}), we see that in this particular case the duration of the pulse
also determines the duration of the interaction 
between the qubit and the environment. 

Whenever (\ref{Hint}) applies, the evolution of the field $\phi$ from a known
initial state is completely determined by the
current $J(\x, t)$, i.e. the switching history. In other words, the gate in this
case works as an antenna, producing a
definite ``radiation'' state of $\phi$. (Again, this ``radiation'' does not have
to be an electromagnetic wave, but can be any kind of propagating excitation.)
Decoherence can be associated with the probability to emit or absorb a nonzero number 
of quanta of $\phi$.

Typically, the initial state of $\phi$ is taken to be a thermal state, with
probabilities of different energy levels given by the canonical distribution
at some temperature $T$. In this paper, we want to deviate from this practice
and consider a microcanonical initial state, in which the oscillators are 
constrained to have their total energy equal to some $E$.\footnote{
The current $J$ will be set to zero at the initial moment, which can always 
be done by a time-independent redefinition of $\phi$. This makes choosing 
a microcanonical initial state for the bath equivalent to choosing it for 
the entire qubit+bath system.}
There are several 
reasons why we think that this problem is interesting and potentially important 
for analysis of various qubit designs.

First, in thermodynamics we are accustomed to canonical and microcanonical 
ensembles being essentially equivalent in the macroscopic limit. It is
interesting to see if, and to what accuracy, the same applies to calculations
of quantum coherence, which is an intrinsically time-dependent quantity.

Second, some of the environments important for current qubit designs are 
in fact comprised by relatively few degrees of freedom. Consider, for example, 
the swap gate described by eq. (\ref{Hdot}), and suppose that the pulse of
voltage is delivered to the gate via a transmission line. Suppose further
that the line is open at one end (where it attaches to the gate) and closed
at the other (and the pulse is obtained, say, through inductive coupling of 
the line to some control circuit). For a line of length $L$, the number of modes
significantly populated at temperature $T$ is of order
\be
N_{\rm eff} \sim \frac{k_B T L}{\pi \hbar c} \; .
\label{N}
\ee
For $L=1$ m and $T=0.1$ K, we obtain $N_{\rm eff}\sim 10$. 
At this point, it is actually not obvious that this $N_{\rm eff}$, i.e. the
number of populated modes, is what controls the transition to the thermodynamic
limit. However, later in the paper we show that this is indeed the case.

Now, although an ensemble of identical oscillators baths contained in
identical experimental apparata may be well described by a thermal density
matrix, in each individual experiment the bath initially has nearly 
fixed energy, with some broadening due to the initial state's
preparation. If the broadening is larger than the level spacing of the bath
but smaller than the typical energy fluctuation in a thermal state,
the microcanonical initial state is a better approximation than the canonical
one. Because the level spacing decreases exponentially with the size of the 
bath, while relative thermal fluctuations only go as inverse square 
root of $N_{\rm eff}$, we expect that such a situation will in fact be 
typical for 
relatively small, ``mesoscopic'' environments. In this case, one may wonder 
how much the microcanonical decoherence, induced by the interaction of the qubit
with the bath, differs from the thermal (canonical) 
result. In particular, one of the main goals of our project was to see
if such a small environment can cause any significant decoherence at
all.

Finally, because decoherence is associated with the response of the environment 
to changes in the system, one may expect that anomalously 
large deviations from the thermal 
result will occur when fluctuations in the environment are large and the 
relaxation is slow, e.g. near a critical point.
Our calculation lends some supports to this idea.

Our main results are as follows. (i) If, for a microcanonical state of energy $E$,
we formally define temperature $T$ by the usual thermodynamic formula, then
the expansion parameter that controls the difference between the canonical and
microcanonical results for decoherence is $1/N_{\rm eff}$, where by definition
$N_{\rm eff} = E/ k_B T$. For a one-dimensional transmission line, this is 
of the same order as eq. (\ref{N}). (ii) We consider an expansion of 
microcanonical decoherence
in powers of $1/N_{\rm eff}$ and find that the leading difference between
the canonical and microcanonical results is formally of order $1/N_{\rm eff}$.
We present both a general formula for this correction, applicable for
any nonlinear environment, and an explicit
formula for an environment with Ohmic dissipation. In particular, for the case
of a transmission line we find that already for $N_{\rm eff}\sim 10$ the 
environment causes
significant decoherence, which is practically indistinguishable from the thermal
result.
(iii) The $1/N_{\rm eff}$ correction contains a term proportional to the derivative
of the heat capacity of the environment with respect to the temperature,
$\partial C_V / \partial T$, which becomes singular near
a critical point. Although in a finite system there can be no 
``true'' critical singularity, a finite enhancement of $\partial C_V / \partial T$
remains. It is significant in this respect that our results
apply to the general
case of a fully nonlinear environment, rather than to a collection of
harmonic oscillators, for which no critical phenomena are expected.
A simple application of finite-size scaling shows that the critical
singularity of $\partial C_V / \partial T$ alone cannot completely cancel
the $1/N_{\rm eff}$ suppression factor (although it can reduce the suppression
considerably). However, the correction also depends on the dissipative
coefficient, characterizing the interactions of the qubit with the environment,
and it is ultimately the critical behavior of this coefficient that determines
both the size and the sign of the correction at a critical point.

In summary, while our results are somewhat inconclusive on the critical behavior
of decoherence (due to the lack of understanding of the critical behavior
of the dissipative coefficient), we obtain a clear demonstration that away
from criticality even a relatively compact, ``mesoscopic'' bath of oscillators,
initially in a microcanonical state,
induces decoherence of practically the same magnitude as a truly macroscopic,
thermal environment.

In the course of the evolution, the qubit and the bath exchange energy,
as described by the interaction Hamiltonian (\ref{Hint}). 
Moreover, because the qubit itself is controlled by some external
means, even the compound qubit+bath system is not strictly isolated (except at
times before and after the switching). We assume, however, that there is no
additional, ``direct'' interaction of the bath with the outside world.
This seems to us a reasonable assumption, since in most cases one will want
to isolate the qubit and its immediate surroundings from the larger 
room-temperature environment.

Our work, then, has some elements
in common with the earlier work of Jensen and  Shankar \cite{JS} on a strictly 
isolated small system. These authors have observed statistical
behavior in a numerical solution of the Schr\"{o}dinger equation for
seven interacting spins. In particular, they have 
found that the distribution of probabilities for one spin in their system
closely resembles the canonical distribution expected if the full spin chain 
were in a microcanonical state.
In the present context, the selected spin plays the role of the qubit (albeit
not subject to any external control), while the remaining spins play the role of 
the bath.

Apart from the question of perfect versus imperfect isolation of the compound
qubit+bath system, the main difference between the work of ref. \cite{JS}
and ours is that \cite{JS} compares results for a pure initial state
to those for a microcanonical ensemble, for a compound system, while we are 
interested in comparing results between canonical and microcanonical initial 
states for such a system.
As we will see, that latter comparison can be done, for 
a rather general case, without a recourse to numerical integrations. Instead, 
our calculation makes use of a steepest-descent evaluation of an integral relating
the canonical and microcanonical averages, 
the accuracy of this procedure being again controlled by $1/N_{\rm eff}$.

The paper is organized as follows. In Sect. 2 we present the definition of
coherence as a functional of the switching history, the latter being
represented
by the current in (\ref{Hint}). We discuss a suitable form of the current.
Although our main results are not based on a perturbative expansion of
coherence, we pause in Sect. 3 to describe a convenient way to perform such 
an expansion, based on the coherent-state formalism. In Sect. 4,
we compute coherence, as defined in Sect. 2, for a thermal initial state
and recover some familiar expressions. In Sect. 5, we
construct the density matrix for a microcanonical initial state.
In Sect. 6, we compute microcanonical decoherence. Sect. 7 is a conclusion.

In what follows we use the system of units with $\hbar = 1$ and $k_B = 1$.

\section{Definition of coherence}
If we know that at some initial time $t=0$, the environment started out in a
definite quantum state $|\Psi(0)\rangle$, we can define coherence remaining in 
the qubit at arbitrary time $t$ in the following way. Find the final state 
of the environment using the evolution operator $U_J(t,0)$, where the ``current'' 
$J$ represents the switching history of the qubit. Coherence equals the overlap 
of that final with the state that would obtain if no switching took place:
\be
C(t) = \langle \Psi(0) | U_0^{\dagger}(t,0) U_J(t,0) |\Psi(0) \rangle \; .
\label{coh1}
\ee
A decrease of the overlap with time (decoherence) is due to the 
divergence of the evolution histories of the environment corresponding to 
different histories of the qubit. It therefore reflects the measuring influence
that the environment had on the qubit.

An obvious extension of this definition to the case when the
state at $t=0$ is a mixed state with a density matrix $\rho(0)$ is
\be
C(t) = {\rm Tr} \left[ U_J(t,0) \rho(0) U_0^{\dagger}(t,0) \right] \; .
\label{coh2}
\ee

It is convenient to incorporate the moments of time $0$ and $t$ in the definition
of the current. To save notation, we describe  the environment by a single real 
scalar $\phi(\x, t)$ with real-valued oscillator modes $\psi_n$:
\be
\phi(\x, t) = \sum_n \phi_n(t) \psi_n(\x) 
\ee
(generalizations are of course possible).
We assume that each mode  couples to some smooth $J_n(\tau)$, 
which is zero at $\tau<0$,
switches on at $\tau\approx 0$, stays on a plateau until $\tau\approx t$, and then
switches off, see Fig. 1. Thus, the Lagrangian of the field is
\be
L = \sum_n \left\{
\frac{1}{2} \dot{\phi}_n^{2} - \frac{1}{2} \omega_n^{2} \phi_n^{2}
+ J_n \phi_n \right\} + L_{\rm int}[\phi] \; ,
\label{L}
\ee
where $L_{\rm int}$ describes a self-interaction.

\begin{figure}
\leavevmode\epsfxsize=5in \epsfbox{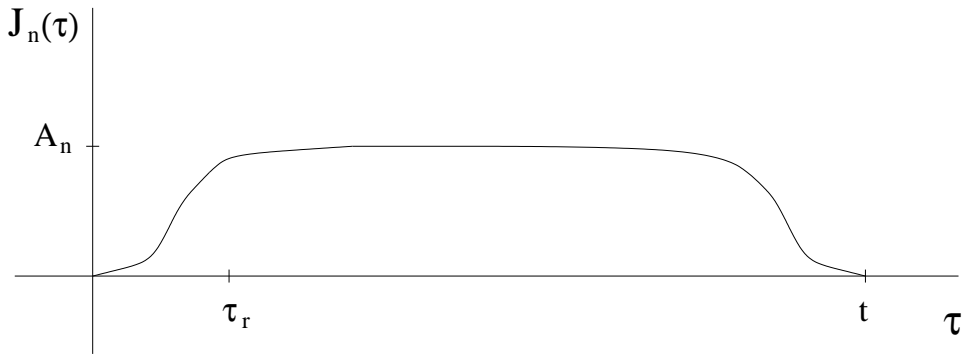}
\vspace*{0.2in}
\caption{Profile of the current $J_n$ representing the switching
history of the qubit.}
\label{fig:Jprofile}
\end{figure}

The above form of the current can describe either of
the following two experimental setups. For the swap gate described by the
Hamiltonian (\ref{Hdot}), the profile shown in Fig. 1
represents a single gate operation: both the initial and final states of
the qubit correspond to $J_n=0$. So, the switching time of the qubit is the entire
time $t$. On the other hand, if one basis state of the qubit corresponds
to $J_n=0$, and another to the plateau value $J_n=A_n$, then the current of Fig. 1
represents two switching operations: from the first state to the second and back.
In this case, the switching time is the ramp time of the current, 
$\tau_r$.\footnote{And coherence defined by (\ref{coh2}) coincides in this case
with what is perhaps a more familiar definition: the value at time $t$ of the
off-diagonal element of the {\em qubit's} density matrix, relative to its value
at $t=0$; $\omega_r = 2\pi / \tau_r$ acts as a frequency cutoff.}

For this form of the current, we can relate $\rho(0)$
to the density matrix at some $T_i < 0$ in the distant past as
\be
\rho(0) = U_J(0,T_i) \rho(T_i) U_0^{\dagger}(0,T_i)
\label{rho0}
\ee
(since at $t<0$ $U_J$ and $U_0$ coincide) and also extend $t$ in (\ref{coh2}) to
some $T_f$ in the distant future. In this way, we obtain
coherence as a functional of $J_n(\tau)$:
\be
C[J] = {\rm Tr} \left[ U_J(T_f,T_i)  \rho(T_i) U_0^{\dagger}(T_f,T_i) \right] \; . 
\label{cohJ} 
\ee
Further, using the environment's $S$-matrix
\be
S_J = \e^{i H_{\rm free} T_f} U_J(T_f, T_i) \e^{-i H_{\rm free} T_i} \; ,
\ee
where $H_{\rm free}$ is the Hamiltonian in the absence of self-interactions
(and interaction with the system), we can rewrite (\ref{cohJ}) as
\be
C[J] = {\rm Tr} \left[ S_J \rho_i S_0^{\dagger}  \right] \; ,
\label{cohS}
\ee
where
\be
\rho_i = \e^{i H_{\rm free} T_i} \rho(T_i) \e^{-i H_{\rm free} T_i} \; .
\label{rhoi}
\ee
Under the usual assumption of adiabatic switching on of the interaction in the
distant past, $\rho_i$ is independent of $T_i$. Thus, specifying it is a convenient
way to impose initial conditions.

Eq. (\ref{cohS}) is the definition of coherence that we use in what follows.
We observe that coherence defined in this way coincides with the generating 
functional of the Green functions corresponding to the state $\rho_i$.
In perturbation theory, it can be computed order by order with the help of
the Schwinger-Keldysh diagram technique.

If we are to have small decoherence, it is natural to assume that the currents
$J_n$ are weak. Then, provided that the field $\phi$ does not have a nontrivial
expectation value, the leading term in $Q[J] = -\ln C[J]$ is bilinear in $J$:
\be
Q[J] = -\frac{i}{2} \sum_{mn} \int dt dt' J_m(t) \Delta_{mn}(t,t') J_n(t') 
+ O(J^3) \; ,
\label{Qgen}
\ee
where $\Delta_{mn}$ is the full (connected) 
Green function of $\phi$ in the state $\rho_i$:
\be
\Delta_{mn}(t,t') = i {\rm Tr} 
\left\{ S_0^{\dagger} T[\phi^I_m(t) \phi^I_n(t') S_0] \rho_i 
\right\} \; ;
\label{greenf}
\ee
$\phi^I$ is the field operator in the interaction representation.
The real part of $Q$, related to the imaginary part of $\Delta$, determines the
exponential suppression of coherence due to switching and can be called the
decoherence exponent.

For specific calculations, we will use the following expression for the Fourier
transform of the current:
\be
\tilde{J}_n(\Omega) = 
\frac{A_n}{i\Omega}\left( \e^{i\Omega t} - 1 \right) \exp(-|\Omega|/2\omega_r) \; ,
\label{specific}
\ee
where $\omega_r = 2\pi / \tau_r$, and $A_n$ are real constants.
In the limit $\omega_r \to \infty$, eq.
(\ref{specific}) becomes the Fourier transform of a
rectangular pulse: $J_n(\tau) = A_n$ for $0<\tau < t$ and zero otherwise.

\section{Perturbative expansion}
Although our main results are not based on a perturbative expansion, 
we pause here to outline a convenient way to carry it out.

As we have seen, coherence naturally acquires an exponential form. So,
it is convenient to compute (\ref{cohS}) in a representation in which
the trace reduces to a saddle-point integral: such integrals produce exponentials
automatically. A good choice is the coherent-state (holomorphic) representation
\cite{Berezin,F&S}, which we now review. 
(For a scattering problem with a large but not macroscopic 
number of particles and microcanonical initial 
conditions, the coherent state representation was used in ref. \cite{periodic}.)

Any state of the environment can be 
represented by an anti-analytical function $\psi(a^*)$ 
of the complex variable $a$ 
labeling the coherent states. Action of an arbitrary operator $\hat{A}$ is
represented by an integral of the form
\be
(\hat{A}\psi) (b^{*}) = \int \frac{da^{*} da}{2\pi i} \e^{- a^{*}a}
A(b^{*}, a) \psi(a^{*}) \; ,
\ee
where $A(b^{*}, a)$ is the kernel of the operator $\hat{A}$ defined by
\be
A(b^{*}, a) = \langle b| \hat{A}|a \rangle \; .
\ee
A product of two operators is represented by the convolution of their
kernels:
\be
(\hat{A_1}\hat{A_2})(b^{*}, b) = \int \frac{da^{*} da}{2\pi i} \e^{-
a^{*}a} A_1(b^{*}, a) A_2(a^{*}, b) \; .
\ee

The $S$-matrix is given by \cite{Berezin,F&S,periodic}
\be
S_J(b^*, a) = \langle b |S_J| a \rangle = \int d\phi_i d\phi_f {\cal D} \phi 
    ~\e^{B_i + B_f + i \int L dt} \; ,
\label{S}
\ee
which contains a functional integral over the field $\phi$ as well as ordinary 
integrals 
over the field's boundary values $\phii$ and $\phif$. 
The boundary terms $B_i$ and $B_f$ read
\ba
B_i & = & \sum_n \left[
- \frac{1}{2} a_n^2  \e^{-2 i \omega_n T_i} - \frac{1}{2} \omega_n \phii^2
+ \sqrt{2 \omega_n} a_n \phii \e^{ - i \omega_n T_i}
\right] \; , \label{Bi} \\
B_f & = & \sum_n \left[
- \frac{1}{2} (b^*_n)^2 \e^{2 i \omega_n T_f} - \frac{1}{2} \omega_n \phif^2 
+ \sqrt{2 \omega_n} b^{*}_n \phif \e^{i \omega_n T_i}
\right] \; . \label{Bf}
\ea
The perturbation expansion for $S_J$ is generated in the usual way via the
relation (see e.g. ref. \cite{F&S})
\be
S_J = \exp[i \int L_{\rm int} (\delta/i\delta J) dt ] S'_J \; ,
\label{Spert}
\ee
where
\be 
S'_J =
\int d\phi_i d\phi_f {\cal D} \phi 
  ~\e^{B_i + B_f + i \int (L_{\rm free} + J\phi) dt} \; ,
\label{S'}
\ee
and $L_{\rm free}$ is the Lagrangian of free oscillators.

The integrals in (\ref{S'}) are
Gaussian and can be evaluated exactly at the corresponding
saddle points. The saddle-point equation for $\phi_n(t)$ is simply the equation
of motion
\be
\ddot{\phi}_n + \omega_n^2 \phi_n = J_n(t) \; ,
\label{eqm}
\ee
while the saddle-point equations for $\phi_i$ and $\phi_f$ supply the boundary
conditions
\ba
\omega_n \phii + i \dot\phii & = & \sqrt{2 \omega_n} 
\e^{- i \omega_n T_i} a_n \; , \label{eqi} \\
\omega_n \phif - i \dot\phif  & = & \sqrt{2 \omega_n} 
\e^{ i \omega_n T_f} b^{*}_n \; . \label{eqf}
\ea
The solution to (\ref{eqm}) with these boundary conditions is
\be
\phi_n(t) = \frac{1}{\sqrt{2 \omega_n}} [ a_n \e^{- i \omega_n t} + 
b^{*}_n \e^{i \omega_n t} ] + \int dt' G_n(t-t') J_n(t')
\label{sp}
\ee
where $G_n$ is the free causal Green function
\be
G_n(t-t') = \frac{i}{2\omega_n} \e^{-i \omega_n |t-t'|}  \; .
\ee
Substituting the saddle-point solution (\ref{sp}) into (\ref{S}), we obtain
(cf. ref. \cite{F&S})
\ba
S'_J(b^*, a) & = & \exp\sum_n \left\{
a_n b^*_n + (i/\sqrt{2 \omega_n}) \int dt J_n(t) 
[a_n \e^{- i \omega_n t} + b^{*}_n \e^{i \omega_n t} ] 
\right. \nonumber \\
 & & \left. + {i\over 2} \int dt dt' J_n(t) G_n(t-t') J_n(t') \right\} \; .
\label{Ssp}
\ea
This can be conveniently rewritten in terms of Fourier 
transforms of $J_n$ and $G_n$, defined as
\ba
\tilde{J}_n(\Omega) & = & 
\int_{-\infty}^{\infty} J_n(\tau) \e^{i \Omega \tau} d\tau \; , 
\label{ftJ} \\
\tilde{G}_n(\Omega) & = & 
\int_{-\infty}^{\infty} G_n(\tau)\e^{i \Omega \tau} d\tau \; .
\label{ftG}
\ea
We obtain
\be
S'_J(b^*, a)  =  \exp\sum_n \left\{
a_n b^*_n
+ \frac{i}{\sqrt{2 \omega_n}} [ \tJ_n^*(\omega_n) a_n + \tJ_n(\omega_n) b^{*}_n ]
+ {i\over 2} \int \frac{d\Omega}{2\pi} \tilde{G}_n(\Omega) 
|\tJ_n(\Omega)|^2 \right\} \; .
\label{ftS}
\ee
This can be used in eq. (\ref{Spert}) to produce a perturbative expansion
for the $S$-matrix.

\section{Thermal decoherence}
Returning to our definition of coherence, eq. (\ref{cohS}), we see that in the
absence of self-interaction we would have $S_0 = 1$ and $S_J = S'_J$, so that
\be
C[J] = {\rm Tr} \left[ S'_J \rho_i  \right] = 
\int \frac{da^{*} da}{2\pi i}\frac{db^{*}db}{2\pi i} 
\e^{- a^{*}a} \e^{-b^{*}b} \rho_i(a^*, b) S'_J(b^*, a) \; . 
\label{Cnint}
\ee
In particular, for a thermal initial state with inverse temperature $\beta$,
\be
\rho_i(a^*, b) = \prod_n \left( 1 - \e^{-\beta\omega_n} \right) 
\exp\left( a^*_n b_n \e^{-\beta\omega_n} \right) \; .
\label{rho_thermal}
\ee
In this case, the integrals in (\ref{Cnint}) are
Gaussian and can be evaluated explicitly. We obtain $C[J] = \exp(-Q[J])$ with
\be
Q[J] = \sum_n \left\{ \frac{|\tilde{J}_n(\omega_n)|^2}{4 \omega_n} 
\left[ 2n_B (\omega_n) + 1 \right] 
- \frac{i}{2} \int \frac{d\Omega}{2\pi} 
|\tilde{J}_n(\Omega)|^2 \Re \tilde {G}_n(\Omega) \right\} \; ,
\label{Qnint}
\ee
where $n_B(\omega) = [\exp(\beta\omega) - 1]^{-1}$ is the Bose distribution.
Eq. (\ref{Qnint}) is the noninteracting limit of the more general 
eq. (\ref{Qgen}).

For an interacting environment, (\ref{rho_thermal}) is still 
the correct initial condition for a thermal state, because the interaction
is assumed absent
in the distant past (and the interacting state is obtained by an adiabatic 
switching on of the interaction, while maintaining the fixed temperature 
$1/\beta$). In the limit of small $J$, we now use eq. (\ref{Qgen}), according
to which the real and imaginary parts of $Q$ are determined, respectively,
by the anti-Hermitean and Hermitean parts of $\tilde{\Delta}_{mn}(\Omega)$ 
(the Fourier transform of $\Delta_{mn}$). The anti-Hermitean part
(which itself is an Hermitean matrix)
\be
{\Delta}''_{mn}(\Omega) = \frac{1}{2i}
\left[ \tilde{\Delta}_{mn}(\Omega) - \tilde{\Delta}^*_{nm}(\Omega) \right]
\ee
can be expressed through the spectral density of the environment $D_{mn}$
in the corresponding channel: at $\omega > 0$,
\be
{\Delta}''_{mn}(\omega) = 
{\Delta}''_{nm}(-\omega) = 
\frac{\pi}{2\omega} D_{mn}(\omega) \coth(\beta \omega / 2)  \; . 
\label{aH}
\ee
$D_{mn}$ includes effects of the self-interaction. 

The real part of $Q$ (the decoherence exponent) becomes
\be
Q_R[J] = \sum_{mn} \int_0^{\infty} \frac{d\omega}{4\omega} 
J^*_m(\omega) D_{mn}(\omega) J_n(\omega) \coth(\beta \omega / 2) \; .
\label{QR}
\ee
The imaginary part is given by
\be
Q_I[J] = -\half \sum_{mn} \int_{-\infty}^{\infty} \frac{d\Omega}{2\pi} 
J^*_m(\Omega) \Delta'_{mn}(\Omega) J_n(\Omega) \; ,
\label{QI}
\ee
where the Hermitean part $\Delta'_{mn}(\Omega)$ can be expressed through
the spectral density $D_{mn}$ via a dispersion relation:
\be
\Delta'_{mn}(\Omega) = {\rm P} \int_0^{\infty} \frac{d\omega}{2\omega}
\left[ \frac{D_{mn}(\omega)}{\omega - \Omega} + 
       \frac{D_{nm}(\omega)}{\omega + \Omega} \right] \; ;
\ee
${\rm P}$ denotes the principal value.

For a current
of the form (\ref{specific}), we can introduce also another kind of 
spectral density,
which takes into account the interaction of $\phi$ with the current:
\be
F(\omega) = \frac{\pi}{2\omega} \sum_{mn} A_m A_n D_{mn}(\omega)  \; .
\label{spec-dens}
\ee
We can now rewrite the decoherence exponent as
\be
Q_R(t) = \frac{1}{\pi} \int_0^{\infty} \frac{d\omega}{\omega^2} F(\omega)
[ 1 - \cos \omega t]
\e^{-\omega / \omega_r} \coth(\beta \omega / 2) \; .
\label{thermal}
\ee
This is, of course, a familiar expression for thermal decoherence, although
it is usually discussed for an environment comprised by harmonic oscillators.
Here, we obtain it for the fully nonlinear case.

For Ohmic dissipation, when
\be
F(\omega) = \eta \omega \; ,
\label{eta}
\ee
the integral in (\ref{thermal}) coincides with an integral computed by Chakravarty 
and Leggett \cite{CL}, so we can use their result to obtain an explicit functional 
form of $Q_R(t)$:
\be
Q_R(t) = \frac{\eta}{2\pi} \ln(1+\omega_{r}^2 t^2) + \frac{\eta}{\pi}
\ln\left[ \frac {\beta}{\pi t} \sinh\frac{\pi t}{\beta} \right] \; .
\label{explicit}
\ee
One should keep in mind, though, that despite this formal similarity, the
macroscopic quantum coherence (MQC) problem, considered in ref. \cite{CL}, 
is different from ours. In the MQC case, transitions between basis states
occur spontaneously, while in a quantum gate they are externally induced.
In particular, the cutoff frequency $\omega_r$ in our case in general 
depends on the switching method.
We also reiterate that in our treatment, the Ohmic form (\ref{eta}) refers to 
a fully interacting environment, rather than to a collection of harmonic
oscillators. So, for example, the dissipative coefficient $\eta$ can now depend 
on temperature.

\section{Microcanonical density matrix}
In the operator language, the microcanonical density matrix for energy $E$
can be written as (cf. ref. \cite{periodic})
\be
\hat{\rho} = {\cal N}^{-1} \delta(\hat{H} - E) = 
\frac{{\cal N}^{-1}}{2\pi}\int_{C} d\xi 
\exp[i\xi(\hat{H} - E)] \; ,
\label{denso-mic}
\ee
where $\hat{H}$ is the Hamiltonian of $\phi$, and ${\cal N}$ is a normalization
factor.
The contour $C$ runs just above the real axis (as shown in Fig. 2 by a dashed
line): a small positive
imaginary part of $\xi$ regulates the contribution of states
with large eigenvalues of $\hat{H}$.

\begin{figure}
\leavevmode\epsfxsize=5in
\epsfbox{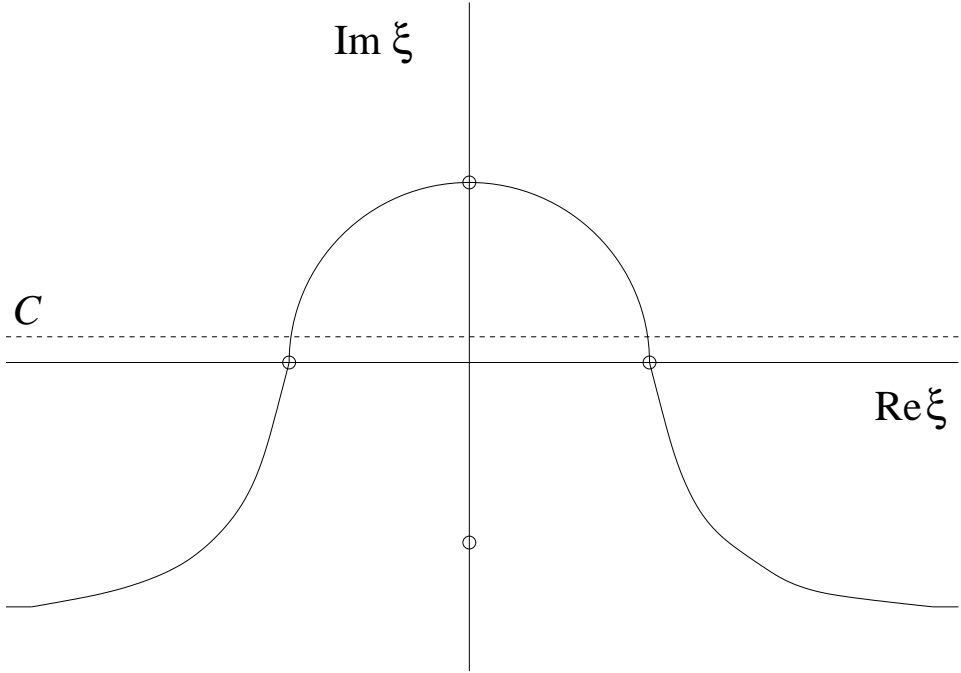}
\vspace*{0.2in}
\caption{The original contour $C$ (dashed line), used in 
the definition (\ref{denso-mic}) of the microcanonical density matrix,
and the deformed contour (solid line), passing through saddle points
(circles). Locations of the four saddle points correspond to an environment 
with an acoustic dispersion law in three spatial dimensions, 
see eq. (\ref{sum1}).}
\label{fig:saddlep}
\end{figure}

In this section, we calculate ${\cal N}$, starting from the normalization
condition
\be
\mbox{Tr}\hat{\rho} = \frac{{\cal N}^{-1}}{2\pi}\int_{C} d\xi 
\e^{-iE\xi} Z(-i\xi) = 1 \; .
\label{norm}
\ee
Here $Z(-i\xi)=\mbox{Tr}\exp(i\xi \hat{H})$ is the thermal partition 
sum analytically continued to a complex inverse temperature $-i\xi$.

The integrand in (\ref{norm}) has a saddle point at $\xi = i \beta$, where
$\beta$ is determined by
\be
E = -\frac{\partial}{\partial \beta} \ln Z(\beta) \; .
\label{Esp}
\ee
Hence, $\beta$ is the inverse temperature related to energy $E$ in the usual
thermodynamic fashion.
The integration contour can
be deformed to pass through the saddle point. Note that it is essential that
the contour was originally defined to run above the real axis, as the
point $\xi =0$ is typically an essential singularity of the integrand in
(\ref{norm}). Calculation by steepest descent in the vicinity of $\xi = i\beta$
gives
\be
{\cal N} = \beta (2\pi/C_V)^{1/2} \exp[E\beta + \ln Z(\beta) ] \; ,
\label{Zsp}
\ee
where $C_V$ is the field's heat capacity:\footnote{Note that using
steepest descent near $\xi = i\beta$ implies 
$C_V > 0$, the usual condition of thermodynamic stability. It is curious
that we have in effect derived this condition without ever referring directly 
to the second law of thermodynamics, the usual source of such inequalities.}
\be
C_V(\beta) = \beta^2 \frac{\partial^2}{\partial \beta^2} \ln Z(\beta) \; .
\label{CV}
\ee

For (\ref{Zsp}) to be a good approximation, two conditions must be satisfied.
First, to use the steepest descent, we must have
\be
 E\beta \gg 1 \; ,
\label{sp-cond}
\ee
and, second, no other saddle point should give a contribution larger than
(\ref{Zsp}). The left-hand side of (\ref{sp-cond}) is our definition of
the effective
number of degrees of freedom, $N_{\rm eff}$. (Recall for comparison that, 
for a collection of noninteracting
classical oscillators at temperature $T$, $E/T$ is precisely
the number of oscillators). So, (\ref{sp-cond}) is the condition that the
environment is relatively macroscopic. 

As for the role of other saddle points,
it has, strictly speaking, to be checked case by case, i.e. for each specific
model of the environment. As an illustration, we include here results for two
simple cases: noninteracting (linear)
environments with an acoustic dispersion law,
\be
\omega_n = v_s k_n \; ,
\label{disp}
\ee
in three and one spatial dimensions. The first case can correspond
for example to phonons, while the second to electromagnetic waves in 
a one-dimensional transmission line (then, $v_s \sim c$).

For linear environments,
\be
\ln Z(-i\xi) = - \sum_n \ln(1 - {\rm e}^{i\omega_n \xi}) \; .
\ee
So, in the first case
\be
\ln Z(-i\xi) = - \const \times \frac{i V}{\xi^3} +
O(L^2) \; ,
\label{sum1}
\ee
where the volume of the tree-dimensional region is denoted by $V$,
and its characteristic linear size by $L$. In the second case,
\be
\ln Z(-i\xi) =  \const \times \frac{iL}{\xi} 
+ O(\ln L) \; ,
\label{sum2}
\ee
where $L$ is the length of the one-dimensional region. 
In both (\ref{sum1}) and (\ref{sum2}), the constants are positive. 
Note that in the case of a transmission line, the field $\phi$
is the ``prepotential'', related
to fluctuations of voltage along the line, $\delta v(x, t)$, via
\be
\delta v = \frac{1}{\sqrt{\cap}} \frac{\partial \phi}{\partial t} \; ,
\label{phi}
\ee
where $\cap$ is the line capacitance per unit length. So, the correlator
of $\delta v$ at coincident $x$ will be Ohmic, with
the ``dissipative coefficient'' 
proportional to the $(\ind/\cap)^{1/2}$ impedance of the line.

We assume that
in both cases the environment is relatively macroscopic, so that the
finite-size corrections indicated in (\ref{sum1}), (\ref{sum2}) are 
negligible. Then, in the case of eq. (\ref{sum1}), the integrand of 
(\ref{norm}) has four saddle-points---at 
$\xi = \beta$, $-\beta$, $i\beta$, and $-i\beta$.
The integration contour can be deformed to pass through the first 
three of these, as shown in Fig. 2, and we find that under condition 
(\ref{sp-cond})
the main contribution to the integral indeed comes from the vicinity of
$\xi = i\beta$. In the case of eq. (\ref{sum2}), there are only two 
saddle points, at $\xi = \pm i\beta$, and only the upper one contributes 
to the integral after deformation of the contour.

\section{Microcanonical Decoherence}
Using the definition (\ref{cohS}) with $\rho_i$ given by 
eq. (\ref{denso-mic}), 
we obtain microcanonical decoherence in the form
\be
C[J] =  {\cal N}^{-1}
 \int_C \frac{d\xi}{2 \pi} \exp \left \{ -i\xi E + \ln Z(-i\xi)
 - Q[\xi, J] \right\} \; ,
\label{Cxi}
\ee
where $Q[\xi, J]$ is the thermal decoherence analytically continued
to a complex inverse temperature equal to $-i \xi$. In the limit of small
$J_n$ (weak decoherence), the real and imaginary parts of the thermal
$Q$ are given by (\ref{QR}), (\ref{QI}).

Let us compare the magnitudes of different terms in the exponent of 
(\ref{Cxi})
on the saddle point $\xi = i\beta$ with $\beta$ determined from (\ref{Esp}).
The first two terms are 
macroscopically enhanced: they are proportional to the effective number of degrees
of freedom
\be
\Neff = E \beta \; .
\label{Neff}
\ee
The third term, $- Q[\xi, J]$, although a sum over $n,m$,
in most cases does not have any macroscopic enhancement, because 
the couplings
$\tilde{J}_n$ scale as $1/\sqrt{\Neff}$, and, while the diagonal entries
of $D_{mn}$ are $O(1)$, most of the off-diagonal entries are $O(1/N_{\rm eff})$.
As a result, to the leading order in $\Neff$, the microcanonical decoherence 
coincides with thermal decoherence at inverse temperature $\beta$, 
while corrections are formally $O(1/N_{\rm eff})$.

Even though $Q[i\beta, J]$ is not enhanced by $\Neff$, in some cases (e.g. for
Ohmic dissipation, cf. (\ref{explicit})) it grows with $t$ (the time for which
$J_n$ is on) and at large $t$ can in principle become a large correction.
However, for applications to qubits, we are interested only in cases
when $Q_R[i\beta, J]$, i.e. decoherence accumulated during time $t$, is much 
smaller than one. 
In these cases, corrections to the thermal result remain formally suppressed by
$1/\Neff$.

Let us calculate the first of these corrections. The exponent of eq. (\ref{Cxi})
can be written as
\be
f[\xi, J] = -iE\xi + \ln Z(-i\xi)  - Q[\xi, J] 
\equiv f_0(\xi) -  Q[\xi, J] \; .
\label{fxi}
\ee
Here, $f_0(\xi) = f[\xi, 0]$ has a saddle point at $\xi = i\beta$, found in the
previous section. The corresponding saddle point of the full $f[\xi, J]$ is 
shifted to
\be
\xi = i \beta + \delta \; ,
\label{xi_sp}
\ee
with a small $\delta$. Treating $Q[\xi, J]$ in (\ref{fxi}) as a perturbation, 
we find 
\be
\delta = \frac{Q'[i\beta, J]}{f''_0(i\beta)} \; ,
\label{delta}
\ee
where primes denote derivatives with respect to $\xi$ (so that $\delta$ is 
in general complex). Note that $f''_0(i\beta) = -C_V / \beta^2$, where
$C_V \propto N_{\rm eff}$ is the heat capacity given by (\ref{CV}).
Therefore, $\delta = O(1/N_{\rm eff})$.

The saddle-point calculation that led to eq. (\ref{Zsp}) in the previous 
section
is now modified in two ways. First, both the extra term in (\ref{fxi})
and the shift of the saddle point contribute to the saddle-point exponent.
Second, they also modify the second derivative of $f$, which determines the
preexponent. As a result, to the leading order in $1/C_V$, we obtain
\be
C(t)  = \left\{ 1 + \half \frac{\partial}{\partial \beta} \left(
\frac{Q_{,\beta}}{f_{0,\beta\beta}} \right) \right\}
\exp\left\{ - Q[i\beta,J] - \half \frac{Q_{,\beta}^2}{f_{0,\beta\beta}} \right\} \; ,
\label{Cmicro}
\ee
where subscripts following commas are used to denote derivatives with
respect to $\beta$. 
The correction to the exponent is always negative, since
$f_{0,\beta\beta} = C_V / \beta^2 > 0$.
Note, however, that although the corrections to both the exponent
and preexponent are of the same order in $1/C_V$, the first is also $O(Q^2)$, while
the second is $O(Q)$. Thus, in the most interesting to us limit of weak decoherence,
the correction to the preexponent is more important. 

In fact, for an interacting
environment, we are not really allowed to keep the correction in the
exponent, since the higher-order terms in $Q$, due to the self-interaction, can
give rise to corrections of the same order, cf. eq. (\ref{Qgen}). We nevertheless
retain this correction in (\ref{Cmicro}) (and in (\ref{Cohmic}) below) because of
the traditional interest in linear environments, for which it is the main
$O(Q^2)$ correction.

As an example, let us take a look at eq. (\ref{Cmicro}) for the case of Ohmic
dissipation. We specialize further to the large-$t$ limit, $t \gg \beta$,
so we can use for $Q_R[i\beta,J]$ the large-$t$ limit of the thermal expression
(\ref{explicit}):
\be
Q_R[i\beta,J] = \frac{\eta t}{\beta} \; ,
\label{large-t}
\ee
while $Q_I$, which is not Bose-enhanced, can be neglected. We find
\be
C(t) = \left\{
1 + \frac{t}{2\beta^2} \frac{\partial}{\partial T} \left( \frac{\eta}{C_V}
\right) \right\}
\exp\left\{ - \frac{\eta t}{\beta} - \frac{\eta^2 t^2}{2\beta^2 C_V} \right\} \; ,
\label{Cohmic}
\ee
where $T=1/\beta$ is the temperature. The correction
to the preexponent, which is the main correction in the limit
\be
1 \ll t / \beta \ll 1/\eta 
\label{limit}
\ee
is negative whenever $C_V/\eta$ is a growing function of $T$. That is the case, 
for example, for linear environments with acoustic dispersion laws, such as 
those considered in the previous section.
However, as we discuss in the conclusion,
there are interesting cases when $\partial C_V / \partial T < 0$, and it is
in principle possible to have a positive correction to coherence.

For the transmission line considered in the previous section, $E \propto T^2$,
so that $C_V = 2 \Neff \propto T$, while $\eta$ is 
$T$-independent. Then, the preexponent in (\ref{Cohmic}) is
equal to $1-\eta t /4\beta \Neff$, which should be compared to 
$1-\eta t /\beta$, the expansion of the exponent in the limit (\ref{limit}).
We see that already for $\Neff = 10$, the correction to the thermal result
is only 2.5\%. 

\section{Conclusion}
Our main result is the calculation of a correction
to the thermal result for decoherence, for a system interacting with a nonlinear 
environment that is initially in a microcanonical (rather 
than canonical) state. 
We expect this result to apply when a qubit interacts mainly with a relatively 
compact, ``mesoscopic'' environment, whose initial spread in energy is smaller
that the typical size of $1/\sqrt{\Neff}$ relative fluctuations characteristic of
a canonical ensemble. 

The correction is given by eq. (\ref{Cmicro}) for
the general case, and by eq. (\ref{Cohmic}) for an Ohmic environment.
We see that the correction is in general of order $1/N_{\rm eff}$ and so
is typically small already for $\Neff \sim 10$, but it
may be enhanced where $\partial C_V / \partial T$ diverges,
i.e. in a proximity of a critical point.

Mathematically,
the correction to the exponent in (\ref{Cmicro}), (\ref{Cohmic}) 
results from the shift in the saddle-point value of $\xi$. According
to (\ref{xi_sp}), such a shift can be interpreted as a change in the effective
temperature of the environment, due to its interaction with the system.
In view of its relation (\ref{QR}) to the anti-Hermitean
part of the full Green 
function, the decoherence exponent $Q_R$ (when it is small) can be interpreted 
as the probability for the system to emit
or absorb an excitation quantum, as a result of the current switching from
$J_n=0$ to $J_n=A_n$. Since in a thermal state the emission is more
probable than the absorption, it is easy to imagine that the change in 
the effective temperature will be positive, leading to an {\em increase}
in decoherence. (The emission probability is 
proportional to $(n_B + 1)$, and the absorption probability to $n_B$; combined,
the two make the $\coth(\beta\omega/2)$ factor in (\ref{QR}).) For example,
for Ohmic dissipation in the $t \gg \beta$ limit, we have
\be
\delta = - \frac{i\eta t}{C_V} \; ,
\ee
which indeed corresponds to an increase in the effective temperature. We recall,
however, that the corresponding increase in decoherence is an $O(Q^2)$ effect, 
subleading in the limit of weak decoherence.

The correction to the preexponent, which is the leading correction in the 
weak-decoherence limit, represents a different phenomenon, namely, a change
in the typical size of fluctuations in the environment as it interacts with 
the system. The enhancement of the correction 
near a critical point reflects the presence of 
large fluctuations at $T= T_c$. Given that, as a condition of 
thermodynamic stability, $C_V > 0$, and that it peaks at $T = T_c$, 
we notice that $\partial C_V /  \partial T < 0$ whenever $T$ is sufficiently 
close to $T_c$ from above. From (\ref{Cmicro}), we see that in this case 
the term containing $\partial C_V /  \partial T$ is positive, 
i.e. it tends to suppress decoherence.

For an environment of a finite size, the singularity of $C_V$ at $T=T_c$ 
appears through $C_V^{-1} \partial C_V /  \partial T$ scaling as some 
positive power of the
total volume. An application of the standard finite-size scaling techniques 
\cite{fs_sc}, together with hyperscaling, gives 
$C_V \propto L^{d+ \alpha/\nu} = L^{2/\nu}$ and
\be
\frac{1}{C_V^2} \left. \frac{\partial C_V}{\partial T} \right|_{T=T_c}
\propto L^{-1/\nu} \; ,
\label{scaling}
\ee
where $L$ is the linear size of the volume, $d$ is the number of dimensions,
$\alpha$ is the specific-heat exponent,
and $\nu$ is the correlation-length one. 
Since $\nu > 0$, (\ref{scaling}) shows that the critical
singularity of $\partial C_V /  \partial T$ cannot completely overcome
the macroscopic suppression (but can reduce it significantly: for comparison,
away from the critical point, the left-hand side of (\ref{scaling}) scales as
$L^{-d}$). However, according to eq. (\ref{Cmicro}), the part of the 
correction that is proportional to $\partial C_V /  \partial T$ is also
proportional to the relevant dissipative coefficient, such as $\eta$ in 
eq. (\ref{Cohmic}). In addition, there is a part of the correction 
containing the derivative of $\eta$. Thus, it is the scaling
of the dissipative coefficient that ultimately determines whether 
the correction to the thermal result can be large enough to be experimentally 
observable.

We note that the critical behavior of the dissipative coefficient determines
also the leading, thermal part of decoherence near a critical point.
As far as we can tell, however, specific results regarding that behavior
are not readily obtainable. In particular, these 
dissipative coefficients are distinct from the usual kinetic coefficients 
introduced in the dynamical theory \cite{dyn_theory}, since they involve
a summation over the modes of
the environment, cf. eq. (\ref{spec-dens}). Our results, then, can 
be taken to underscore the importance of a study of these quantities for
different types of interacting environments.

\acknowledgements
G.S. was supported in part by a grant from Purdue Research Foundation.

\end{document}